\documentclass[aps,prd,twocolumn,showpacs,nofootinbib,amsmath,amssymb,floatfix,superscriptaddress,showkeys]{revtex4}
\usepackage{graphicx}
\usepackage{dcolumn}
\usepackage{bm}
\usepackage{captcont}
\usepackage{xcolor}
\usepackage{threeparttable}
\usepackage{epstopdf}
\usepackage{mathtools}
\usepackage{natbib}
\usepackage{ulem}
\definecolor{blue0}{rgb}{0,0,0.6}
\usepackage[colorlinks,linkcolor=blue0,anchorcolor=blue0,citecolor=blue0,urlcolor=blue0]{hyperref}

\newcommand{\apjl}{Astrophys. J. Lett.}

\newcommand{\mnras}{Mon. Not. R. Astron. Soc.}

\newcommand{\aj}{Astron. J.}

\newcommand{\pasp}{PASP}

\begin{document}

\title{Search for GeV flare coincident with the IceCube neutrino flare}

\author{Yun-Feng Liang}
\affiliation{Key Laboratory of Dark Matter and Space Astronomy, Purple Mountain Observatory, Chinese Academy of Sciences, Nanjing 210034, China}

\author{Hao-Ning He}
\email{hnhe@pmo.ac.cn}
\affiliation{Key Laboratory of Dark Matter and Space Astronomy, Purple Mountain Observatory, Chinese Academy of Sciences, Nanjing 210034, China}
\affiliation{Astrophysical Big Bang Laboratory, RIKEN, Wako, Saitama, Japan}

\author{Neng-Hui Liao}
\affiliation{Key Laboratory of Dark Matter and Space Astronomy, Purple Mountain Observatory, Chinese Academy of Sciences, Nanjing 210034, China}

\author{Yu-Liang Xin}
\affiliation{Key Laboratory of Dark Matter and Space Astronomy, Purple Mountain Observatory, Chinese Academy of Sciences, Nanjing 210034, China}

\author{Qiang Yuan}
\affiliation{Key Laboratory of Dark Matter and Space Astronomy, Purple Mountain Observatory, Chinese Academy of Sciences, Nanjing 210034, China}

\author{Yi-Zhong Fan}
\email{yzfan@pmo.ac.cn}
\affiliation{Key Laboratory of Dark Matter and Space Astronomy, Purple Mountain Observatory, Chinese Academy of Sciences, Nanjing 210034, China}

\date{\today}
 
\begin{abstract}
Recently the IceCube collaboration and 15 other collaborations reported the spatial and temporal coincidence between the neutrino event IceCube-170922A and the radio-TeV activity of the blazar TXS 0506+056. Their further analysis on 9.5 years of IceCube data discovered neutrino flare between September 2014 and March 2015, when TXS 0506+056 is however in ``quiescent"  state. We analyze the Fermi-LAT data in that direction, and find another bright GeV source PKS 0502+049, which is at an angle of $1.2^{\circ}$ from TXS 0506+056, with strong activties during the neutrino flare. No other bright GeV source was detected in the region of interest. 
Though PKS 0502+049 is $1.2^\circ$ separated from TXS 0506+056, it locates within the directional reconstruction uncertainties of 7 neutrinos, out of the 13 neutrino events during the neutrino flare.
Together with the observed high flux of the $\gamma$-ray flare, it may be unreasonable to fully discard the (partial) contribution of PKS 0502+049 to the neutrino flare. The single source assumption used in the neutrino data analysis might need to be revisited.
\end{abstract}


\maketitle

\section{Introduction} \label{sec:intro}

The IceCube collaboration discovered TeV-PeV astrophysical neutrinos in 2013 \citep{IceCube2013}.
Though a neutrino event is found in a spatial coincidence with a GeV-brightening phase of a blazar PKS B1424-418 \citep{Kadler2016BlazarNeutrino}, no associations between neutrinos and point sources with high significance are found until recently.
On 2017 Sep. 22, the IceCube observation detected a track-like neutrino IceCube-170922A, which is found to be associated with the source TXS 0506+056 spatially and temporally, with a significance level of $3\sigma$ \citep{IceCube2018MultiMessenger}. 
The TXS 0506+056 is a blazar active in the optical-TeV energy range. Though its properties is found typical among the bright Fermi-LAT blazars \cite{Liao2018}, it has very strong GeV activities among BL Lacs and is located at a declination toward which the IceCube has highest sensitivity \cite{IceCube2018neutrinoflare}.
It is the first time of finding the association between neutrino and point source with such a high significance. 
It provides evidence for that the AGN jets can accelerate very high energy cosmic rays, and produce neutrinos via the photohadronic interaction or the hadronuclear interaction 
\citep{Protheroe1997Blazar,Dermer2001Blazar,Gao2018blazarneutrino,Murase2018blazarneutrino,Righi2018blazarneutrino,Ansoldi2018blazarneutrino,Keivani2018blazarneutrino,Rodrigures2018blazarneutrino, Liu2018blazarneutrino}.

After the detection of IceCube-170922A, the IceCube collaboration re-analyzed their historical data in the direction of TXS 0506+056 and surprisingly found evidence for a flare of muon-neutrino events between September 2014 and March 2015 \citep{IceCube2018neutrinoflare}. 
The ``puzzle" is the appearance of such a neutrino flare in the quiescent state of both the radio and GeV emission of the blazar TXS 0506+056, though the GeV spectrum got hardened \citep{Padovani2018}. Therefore it is helpful to check whether there are any other ``nearby" flaring GeV sources. For such a purpose we analyze the Fermi-LAT data in the direction of TXS 0506+056, and find that the other bright blazar PKS 0502+049 (also known as 3FGL J0505.3+0459), separated by an angular distance of $1.2^{\circ}$ \citep{Johnston1995} from TXS 0506+056,
had strong GeV flares during the neutrino flare. 
No other flaring GeV source was detected in the neutrino flare phase. In this work we report the results of our data analysis and discuss their implication.    

\section{Fermi-LAT data analysis and detection of GeV flares from PKS 0502+049}\label{sec:Data}
In this work we analyze the publicly-available Pass 8 data of the Large Area Telescope onboard the Fermi space gamma-ray telescope  \citep{Atwood2009,Atwood2013}. 
We perform a standard Fermi-LAT unbinned likelihood analysis\footnote{\url{https://fermi.gsfc.nasa.gov/ssc/data/analysis/scitools/likelihood_tutorial.html}} 
to derive the light curves of TXS 0506+056 and PKS 0502+049.
The gamma-ray data in the energy range of 100 MeV to 500 GeV and in the period of 2010-05-01 to 2018-05-01 (MET 294364802 - 546825605, i.e., MJD 55317 - 58239) are used in the analysis. The start point of the time period is roughly the beginning of IC79 phase of IceCube \citep{IceCube2018neutrinoflare}.
We select gamma rays within a region of interest (ROI) of $10^\circ$ circle centered on the PKS 0502+049 and use a model containing both Fermi-LAT FL8Y
\footnote{\url{https://fermi.gsfc.nasa.gov/ssc/data/access/lat/fl8y/}} sources and diffuse backgrounds ({\tt gll\_iem\_v06.fits} and
{\tt iso\_P8R2\_SOURCE\_V6\_v06.txt}) to fit the data. 
To determine the parameters of the background sources, we first perform a global fitting in which all spectral parameters of the sources located within the ROI are free parameters.
When deriving the light curves, in each time bin, the spectral index parameters are fixed to the values determined in the global fitting, and only prefactors are free to vary. But for PKS 0502+049 and TXS 0506+056, the two sources that we are mainly interested in, if their TS values are greater than 25 in that bin, the spectral index parameters are also set to be free parameters.

\begin{figure}
\centering
\includegraphics[width=0.99\columnwidth]{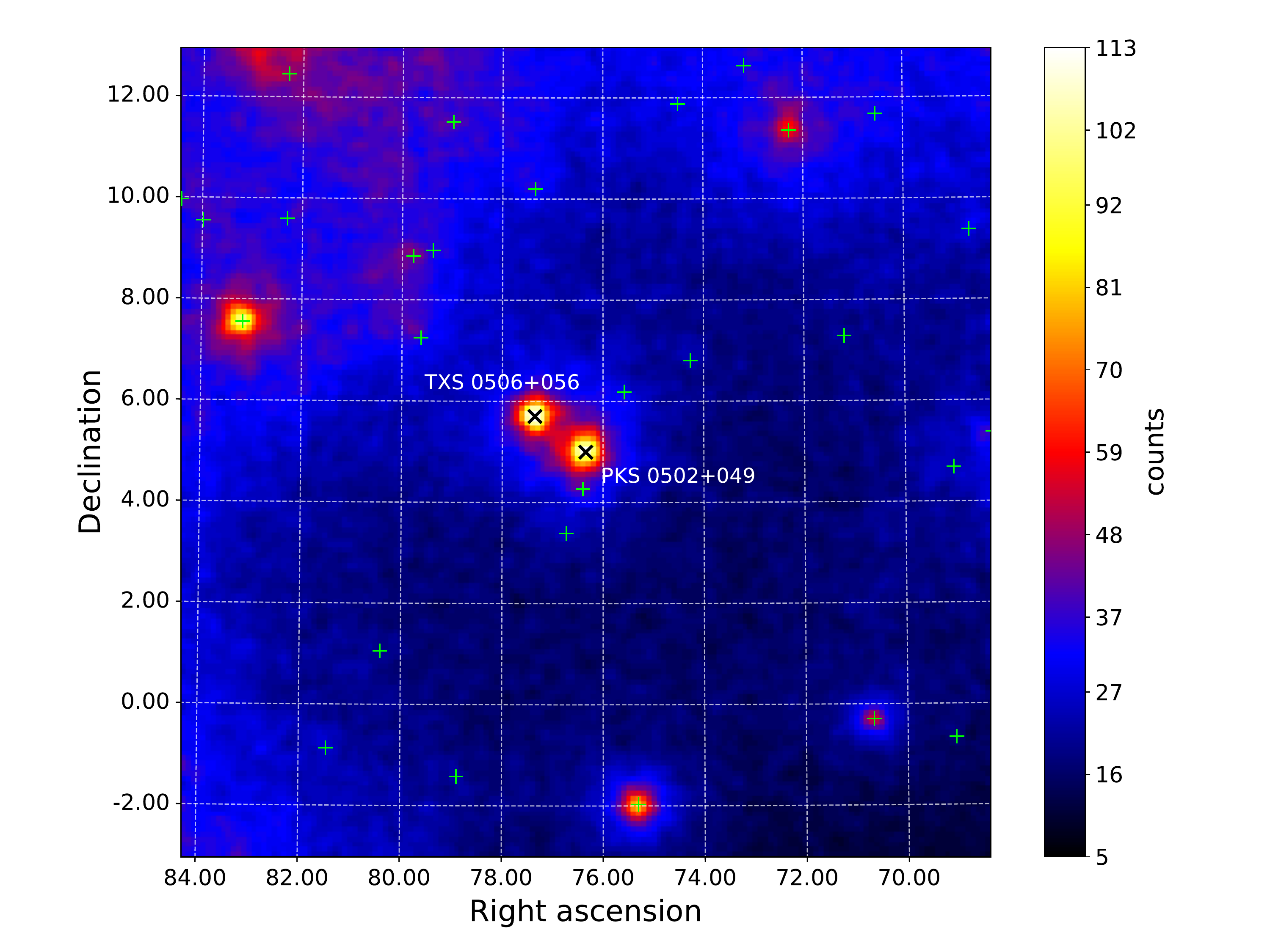}
\caption{$16^\circ\times16^\circ$ counts map centered on the position of PKS 0502+049. The counts map has been smoothed with a Gaussian kernel of $0.3^\circ$. The green plus points are background point sources located within the map. And the two black cross points are for PKS 0502+049 and TXS 0506+056, respectively.
}
\label{fig:cmap}
\end{figure}

\begin{figure}[!b]
\centering
\includegraphics[width=0.99\columnwidth]{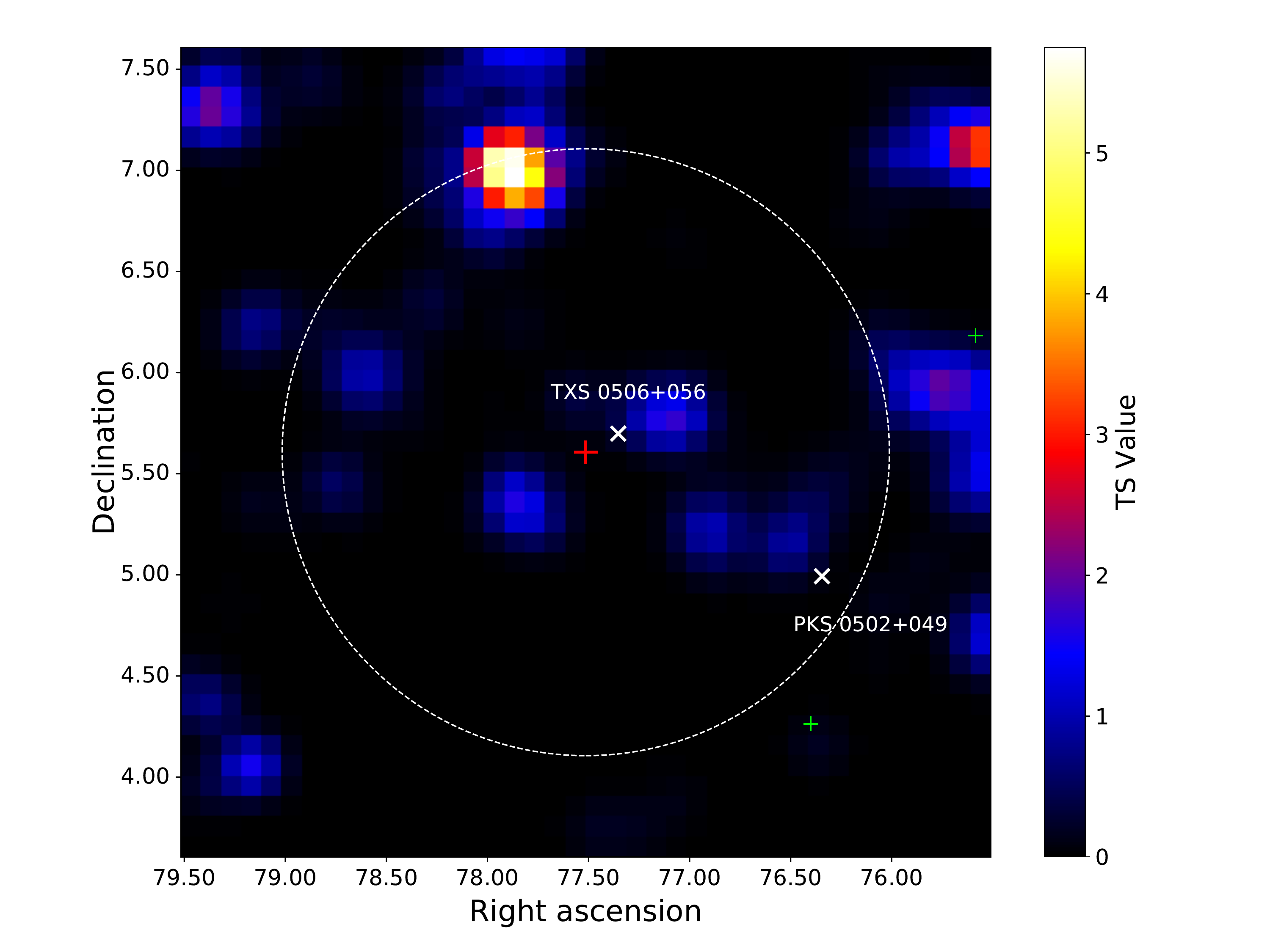}
\caption{The $4^\circ{\times}4^\circ$ residual TS map in the neutrino flare period with TXS 0506+056, PKS 0502+049 and all background sources subtracted. No other gamma-ray source with ${\rm TS}>25$ is seen surrounding TXS 0506+056. The ``cross" in red represents the maximum $-\log_{10}(P)$ position of the neutrino flare signal \citep{IceCube2018neutrinoflare}. 
The white dashed circle encompasses a $1.5^\circ$ region.
}
\label{fig:tsmap}
\end{figure}

In Fig.\ref{fig:cmap}, we show the 100 MeV$-$500 GeV counts map in the direction of TXS 0506+056. There is a nearby bright GeV source PKS 0502+049, located at right ascension (RA) $76.35^\circ$, declination (Dec) $+5.00^\circ$ (J2000 equinox) \citep{Gaia2018}, and redshift $z=0.954$ \citep{Drinkwater1997}, with a spatial separation of $1.2^{\circ}$ to TXS 0506+056. 
Besides these two point sources (i.e., TXS 0506+056 and PKS 0502+049), within the 158-days time window of the neutrino flare\citep{IceCube2018neutrinoflare}, there is no evidence for any other reliable GeV source within the nearby region around the optimal position of the neutrino flare (see Figure \ref{fig:tsmap}).

\begin{figure*}[!htb]
\centering
\includegraphics[width=0.45\textwidth]{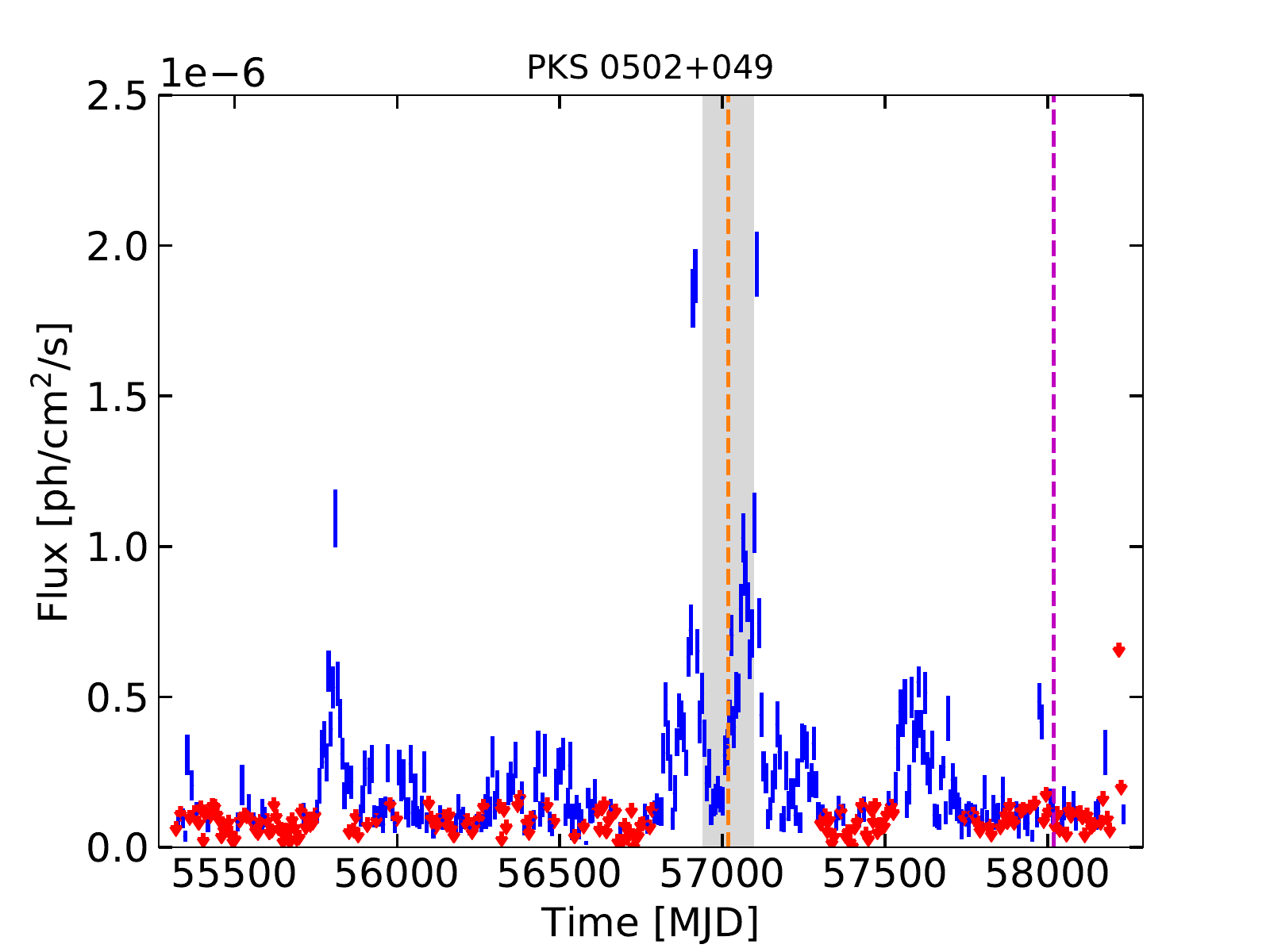}
\includegraphics[width=0.45\textwidth]{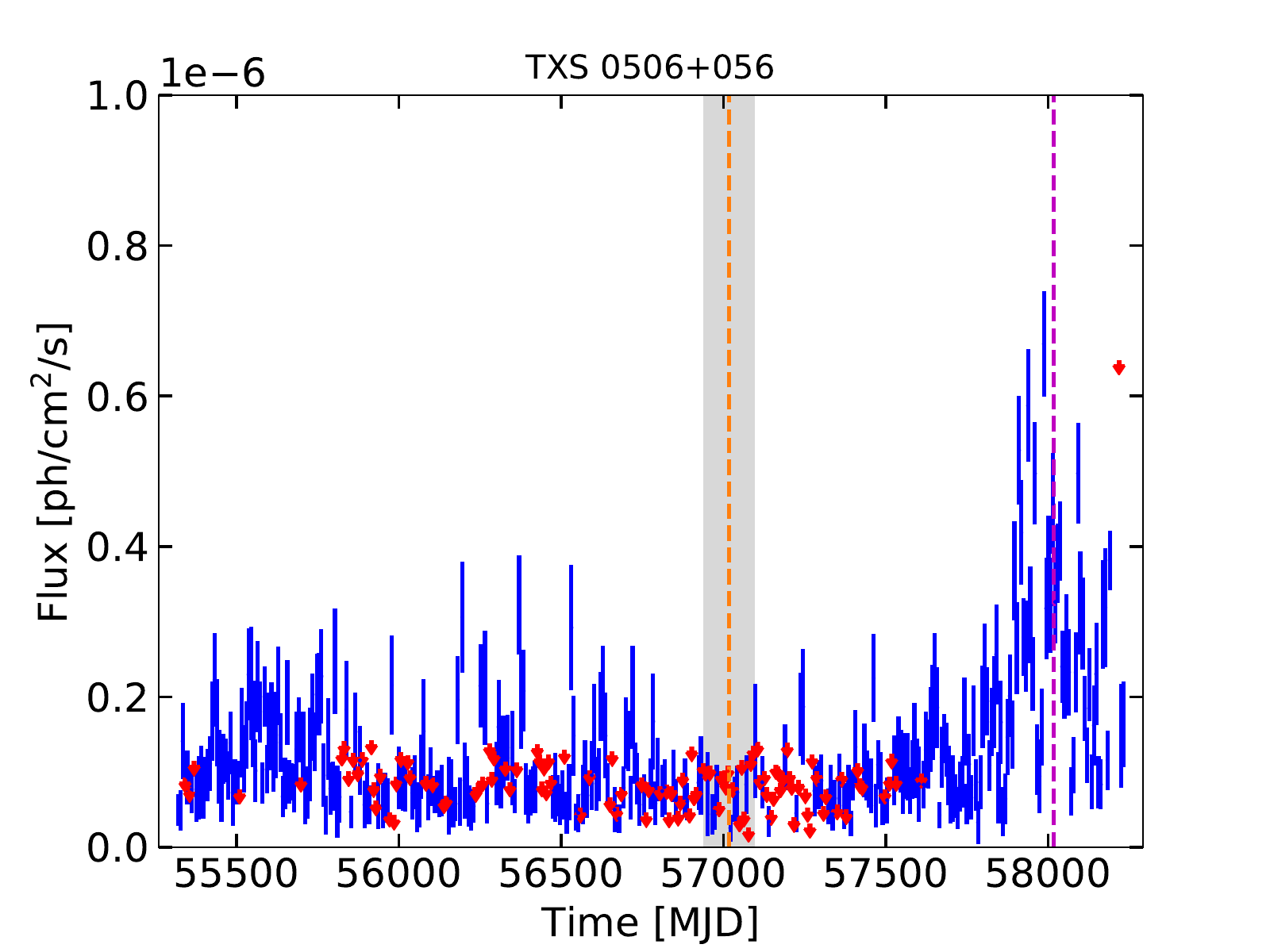}
\caption{The week-bin light curves of PKS 0502+049 (left) and TXS 0506+056 (right). The magenta dashed line represents the arrival time of Icecube 170922. The orange line denotes the center of the box-shaped time window for the neutrino emission reported in \cite{IceCube2018neutrinoflare}, with the whole 158-day time window centered at MJD 57017 shown as shaded area.
}
\label{fig:weeklc}
\end{figure*}

More intriguingly, PKS 0502+049 had strong GeV flares around the time window of the IceCube neutrino flare, while TXS 0506+056 was in a quiescent state (see Fig.\ref{fig:weeklc}; please see also 
http://www.astro.caltech.edu/ovroblazars/data.php for the low radio emission of TXS 0506+056 during the neutrino flare). 
The two major active phases of PKS 0502+049 appeared in the periods of MJD 56860 - 56960 and MJD 57010 - 57120, which is partly overlapped with the 158-days box-shaped time window of IceCube neutrino flare from MJD 56937 to MJD 57096 \citep{IceCube2018neutrinoflare}, as is shown in the shaded area in Figure \ref{fig:weeklc}.
We note that \citet{Padovani2018} has also compared the IceCube neutrino flare with the activity of the PKS 0502+049. However, they demonstrated that the neutrino flare is right between the two gamma-ray flare phases but not in coincidence with them. The reason is that they adopted the 110-day Gaussian-shaped time window of neutrino flare, which is shorter than the box-shaped time window.

\begin{figure*}[htb]
\centering
\includegraphics[width=0.45\textwidth]{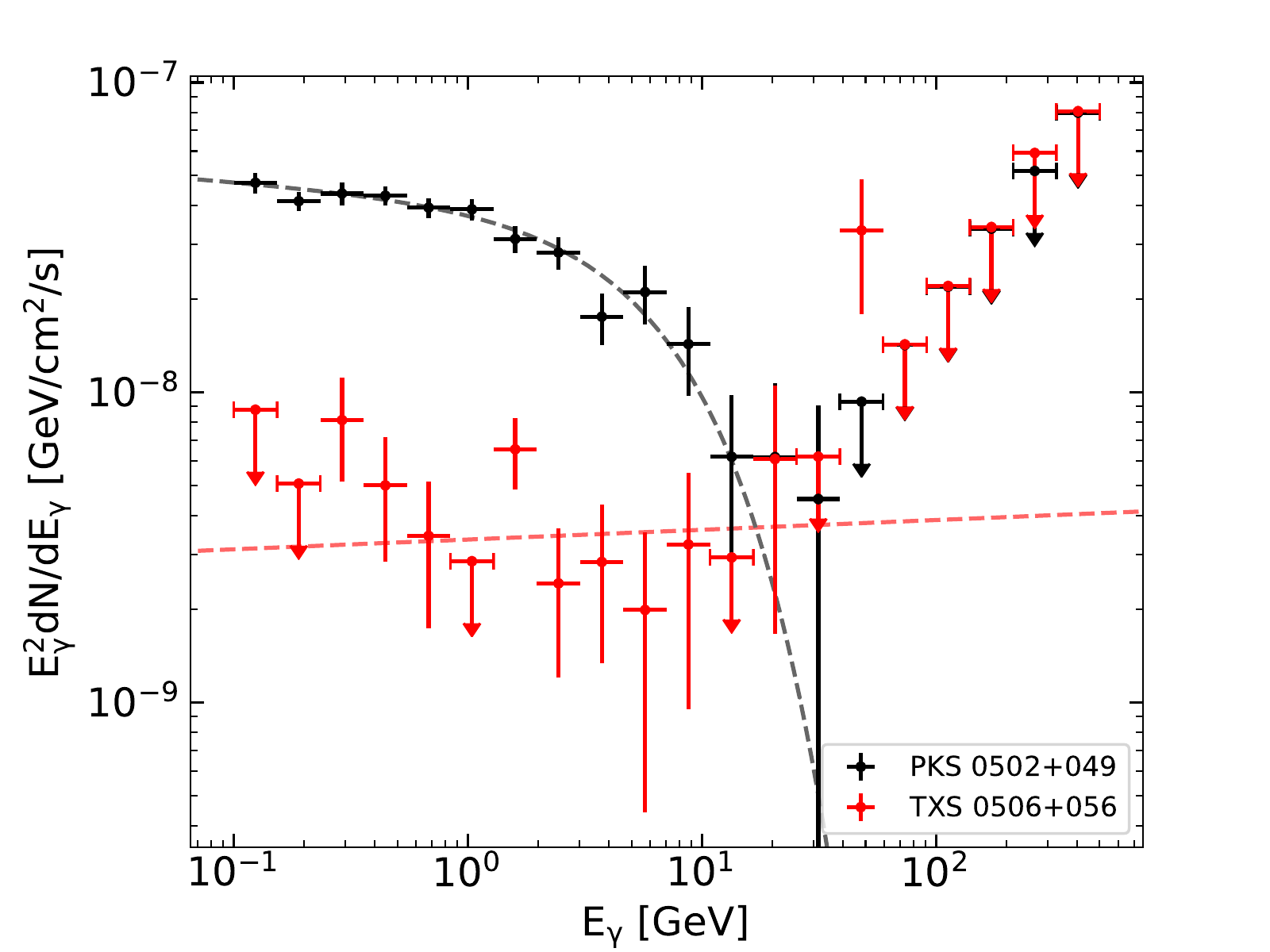}
\includegraphics[width=0.45\textwidth]{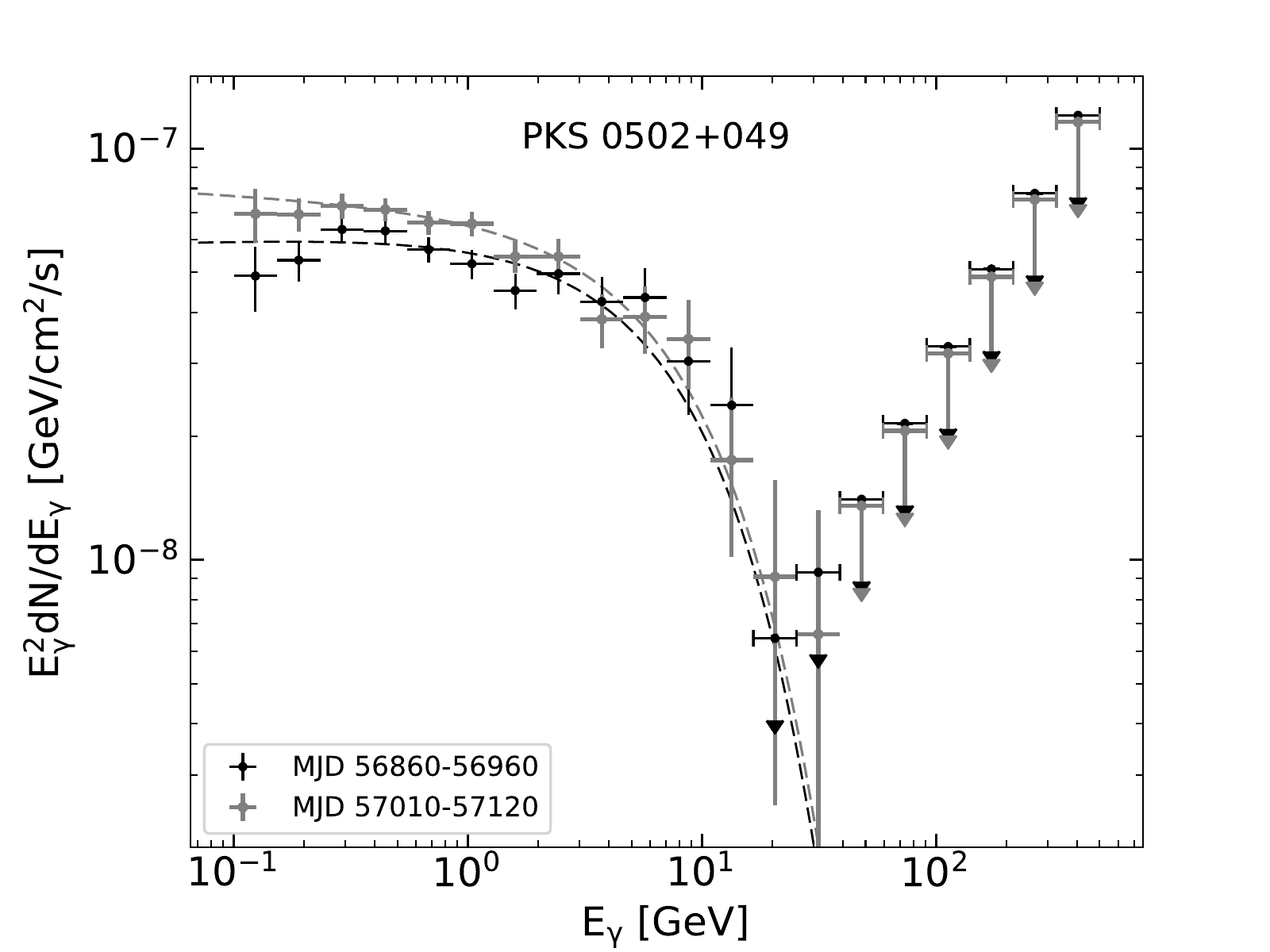}
\caption{Left panel: The spectral energy distributions for the sources of TXS 0506+056 (red) and PKS 0502+049 (black) in the neutrino flare phase (MJD 56937.81 to MJD 57096.21). Upper limit is presented if the TS value in that bin is lower than 4. Right panel: The SEDs of PKS 0502+049 in two major flare phases (MJD 56860 - 56960 and MJD 57010 - 57120) around the neutrino flare. In the left panel, the black dashed line is the best fit ECPL spectrum for PKS 0502+049, with the power law index $p=2.05 \pm 0.04$ and cutoff energy $E_{\rm cut}=7.4 \pm 1.8\,{\rm GeV}$. The red line is the best fit power law spectrum for TXS 0506+056 with index $p=1.97 \pm 0.12$. The ECPL function fit to the two flare phases in the right panel results in $p=1.98 \pm 0.04$ ($2.03 \pm 0.03$) and $E_{\rm cut}=8.6 \pm 2.0\,(9.0 \pm 2.0)\,{\rm GeV}$ for the first (second) flare.
}
\label{fig:spec}
\end{figure*}

In the left panel of Figure \ref{fig:spec}, we compare the spectra of TXS 0506+056 and PKS 0502+049, averaged over the shaded time window shown in Figure \ref{fig:weeklc}, i.e., the period of the neutrino flare.
The right panel shows the spectra of PKS 0502+049 in its two major flare phases of MJD 56860 - 56960 and MJD 57010 - 57120.
The GeV flux of PKS 0502+049 is significantly higher than that of TXS 0506+056 at the energies lower than 10 GeV,
while the spectrum becomes softer at the energies higher than 10 GeV.
The spectrum of PKS 0502+049 can be fitted via a power law spectrum with an exponential cutoff (ECPL), i.e., $\frac{dN_\gamma}{dE_\gamma}\propto E_\gamma^{-p}{\rm exp}(-E_\gamma/E_{\gamma,\rm cut})$, with the index of $p=2.05\pm 0.04$ and the cutoff energy of $E_{\rm cut}=7.4\pm 1.8~\rm GeV$. 
The high energy cutoff of the observed gamma-ray spectrum may be 
due to the attenuation of the high energy photons via the pair production process of interacting with the UV-optical photons from the broad line regions (BLRs) \citep{Poutanen2010,Stern2014nH,he2018}.
The gamma-ray spectrum of TXS 0506+056 becomes harder at the high energy end, the spectrum of which is fitted with a single power law with index of $1.97\pm0.12$.

\section{Discussions}
The IceCube detected a neutrino flare around the location of the blazar TXS 0506+056, which is the object associated with the neutrino event IceCube-170922A.
Via searching for the GeV activity around the location of the neutrino flare, we find that TXS 0506+056 is in a quiescent state, while another blazar, PKS 0502+049, is active around the time of the neutrino flare.
PKS 0502+049 is separated from the optimal position of the neutrino flare by an angle of $1.2^\circ$. 
Though the median angular resolution of the muon neutrino events is about $0.5^\circ$ at energies around $\sim 30$ TeV \citep{IceCube20177years,IceCube2018neutrinoflare},
the directional reconstruction uncertainty for each individual event can range from $\sim 0.2^\circ$ to $>2^\circ$.
IceCube has released the information of neutrino events around the direction of TXS 0506+056, including the time, the reconstructed direction and the directional reconstruction uncertainty.
We denote the directional reconstruction uncertainty $\sigma_{i}$ for the $i$-th neutrino event.
Therefore, we can calculate the angular separation between PKS 0502+049 and the reconstructed direction of the $i$-th neutrino event,
marked as $\sigma_{i,\rm sP}$.
We define those neutrino events with $\sigma_{i,\rm sP}\leq \sigma_{i}$ to be neutrino events that are potentially associated with PKS 0502+049.
Eight neutrino events could be associated with PKS 0502+049. 
Intriguingly, among the 13 neutrino events sugegsted to be related to TXS 0506+056, 7 are also spatially consistent with PKS 0502+049.
Therefore, we cannot exclude the possibility that PKS 0502+049 contributed (partly) to the neutrino flare from 2014 to 2015.
As a consequence, the assumption that the neutrino flare was from a single source may need to be revisited. 
{A time dependent analysis by IceCube, assuming two possible sources TXS 0506+056 and PKS 0502+049, is encouraged.}

In hadronuclear interactions,  the ratio between the flux of $\gamma$-rays to muon neutrinos is $N_{\gamma}:N_{\nu_\mu}\sim 2:1$ \citep{Kelner2006pp}, considering the equipartition among the three neutrino flavors after their oscillations during propagation.
Therefore, if assuming the gamma-ray emission from the two blazars are from pion decay,
we can estimate the counts of neutrinos via extrapolating their gamma-ray spectra to high energy.
If assuming that the softness of the gamma-ray spectrum at $E_\gamma>10~{\rm GeV}$ of PKS 0502+049 is due to the attenuation of low energy photons,  
we can estimate the muon neutrino flux $E_{\nu_\mu}^2{dN_{\nu_\mu}}/{dE_{\nu_\mu}}$ at $100~{\rm TeV}$ as $\sim2.2\times10^{-8}{\rm ~GeV ~cm^{-2} ~s^{-1}}$ and $\sim5.8\times10^{-9}{\rm ~GeV ~cm^{-2} ~s^{-1}}$, 
via adopting the fitted power law index of $2.05$ and $1.97$, for the spectra of PKS 0502+049 and TXS 0506+056, respectively.
Approximating the effective area of IceCube as $A_{\rm eff} \sim 100~{\rm m^2}$ for $\sim$100 TeV neutrinos \citep{IceCube2018neutrinoflare}, the number of muon neutrino events detectable by IceCube during the neutrino flare era is estimated to be $\sim 3$ and $0.4$, for PKS 0502+049 and TXS 0506+056, respectively. 
Based on the above assumptions, by extrapolating the observed GeV gamma-rays, we find that PKS 0502+049 can contribute a fraction of the neutrino flare, while the expected neutrinos from TXS 0506+056 may be much less than the observed counts.
One possibility is that the gamma-ray emission from TXS 0506+056 are attenuated by a fraction of $\sim 70\%$.
Another possibility is that the gamma-rays produced via the hadronuclear interaction contribute to higher energy band, 
while the GeV gamma-ray emission of TXS 0506+056 is not from the hadronuclear interaction \citep{IceCube2018neutrinoflare}.

\begin{table}[!t]
\centering
\caption{\label{table} Features of the blazars TXS 0506+056 and PKS 0502+049.}
\begin{tabular}{ccc}
\hline
\hline
Source Name&TXS 0506+056&PKS 0502+049\\
\hline
Redshift&0.34&0.95\\
Classification&BL Lac&FSRQ \\ 
Jet &Yes & Yes\\   
BLRs & -- & Yes \\
$M_{\rm BH}~[10^8 M_{\odot}]$&--&7.5\\
$L_{\rm Edd}~[10^{47}{\rm erg~ s^{-1}}]$&--&0.95\\
Activity Status&Quiet&Active\\
$L_{\gamma}~[10^{48}{\rm erg~s^{-1}}]$&0.02&1.3\\
$N_{\nu_\mu}^{\rm obs}$&13&8\\
\hline
\end{tabular}
\begin{tablenotes}
\item $\ast$ $N_{\nu_\mu}^{\rm obs}$ is the number of observed muon neutrinos that are associated with each source during the 158-day time window of the neutrino flare.
\end{tablenotes}
\end{table}

In Table \ref{table}, we summarize the features of TXS 0506+056 and PKS 0502+049.
These two sources are both blazars with jets pointing towards us.  
TXS 0506+056 is classified to be a BL Lac object without observation of BLRs \citep{Paiano2018redshift},
while PKS 0502+049 is classified to  be a flat spectrum radio quasar (FSRQ) displaying BLRs.
The flux of PKS 0502+049 at $\sim$ GeV is about one order of magnitude higher than that of TXS 0506+056.
Moreover, PKS 0502+049 is about 3 times more distant than TXS 0506+056.
Therefore, the gamma-ray luminosity of PKS 0502+049 is about two orders of magnitude higher than that of TXS 0506+056.

Since The observed gamma-ray luminosity of TXS 0506+056 is smaller than the luminosity of the neutrino flare \citep{IceCube2018neutrinoflare},
to interpret TXS 0506+056 as the source of the neutrino flare, one has to assume that the GeV gamma-rays are highly absorbed or the energies of gamma-rays from pion decay are higher than 10 GeV. 
If the neutrino flare is associated with the TXS 0506+056, it may be reasonable to speculate a hadronuclear interaction origin of the IceCube neutrino flare, since its radio, optical\citep{Kochanek2017ASAS-SN} and GeV emission are all in low state.  
On the other hand, since PKS 0502+049 is bright in GeV band, its contribution to neutrinos may be non-ignorable.  
The observable BLRs of PKS 0502+056 provide targets for the haronuclear interaction. 
A jet-cloud interaction model is able to interpret the neutrino flare and the gamma-ray flare from PKS 0502+049 simultaneously\citep{he2018}.

\section{summary}
The sources of astrophysical TeV-PeV neutrinos and ultra-high energy cosmic rays are one of the key issues to be solved in this community. Thanks to the successful performance of IceCube, Fermi-LAT, MAGIC, and the quick follow-up observations in optical, radio and X-ray by the other collaborators, significant progresses have been made recently \citep{IceCube2018MultiMessenger}. 
The most encouraging news so far is the spatial and temporal coincidence between the neutrino event IceCube-170922A and the radio-TeV active blazar TXS 0506+056. 

Intriguingly, the IceCube collaboration analyzed the historical data in the direction of TXS 0506+056 and found evidence for a neutrino flare ranging from September 2014 to March 2015 \citep{IceCube2018neutrinoflare}. TXS 0506+056, however, was in quiescent state during the neutrino burst. To search for possible ``nearby" GeV outburst in such a period, we have analyzed the Fermi-LAT data in the direction of TXS 0506+056.

We found a bright source PKS 0502+049, which had strong GeV flares around the neutrino flare phase.
It locates $\sim 1.2^\circ$ away from TXS 0506+056, 
but within the directional reconstruction uncertainties of 7 neutrino events of the neutrino flare.
Its contribution to the neutrino flare may be non-ignorable.
Therefore, the single-source assumption in \citep{IceCube2018neutrinoflare} may be invalid.
The time-integrated \citep{Braun2008pointsourceneutrino} and time-dependent \citep{Braun2010Time-dependent} searches in IceCube historical data as discussed in \citep{IceCube2015timedependent,IceCube2017timeintegrated,IceCube2018neutrinoflare} in the direction of PKS 0502+049 are encouraged.

\acknowledgments
We thank the useful discussion with Ziqing Xia, Ruoyu Liu and Nagataki Shigehiro.
This work was supported in part by NSFC under grants of No.11525313 (i.e., Funds for Distinguished Young Scholars) and No.11703093.
H.N.H. is supported by National Natural Science of China under grant 11303098, and
the Special Postdoctoral Researchers (SPDR) Program in RIKEN.

\bibliographystyle{apsrev4-1-lyf}
\bibliography{neutrino201808}

\begin{thebibliography}{31}%
\makeatletter
\providecommand \@ifxundefined [1]{%
 \@ifx{#1\undefined}
}%
\providecommand \@ifnum [1]{%
 \ifnum #1\expandafter \@firstoftwo
 \else \expandafter \@secondoftwo
 \fi
}%
\providecommand \@ifx [1]{%
 \ifx #1\expandafter \@firstoftwo
 \else \expandafter \@secondoftwo
 \fi
}%
\providecommand \natexlab [1]{#1}%
\providecommand \enquote  [1]{``#1''}%
\providecommand \bibnamefont  [1]{#1}%
\providecommand \bibfnamefont [1]{#1}%
\providecommand \citenamefont [1]{#1}%
\providecommand \href@noop [0]{\@secondoftwo}%
\providecommand \href [0]{\begingroup \@sanitize@url \@href}%
\providecommand \@href[1]{\@@startlink{#1}\@@href}%
\providecommand \@@href[1]{\endgroup#1\@@endlink}%
\providecommand \@sanitize@url [0]{\catcode `\\12\catcode `\$12\catcode
  `\&12\catcode `\#12\catcode `\^12\catcode `\_12\catcode `\%12\relax}%
\providecommand \@@startlink[1]{}%
\providecommand \@@endlink[0]{}%
\providecommand \url  [0]{\begingroup\@sanitize@url \@url }%
\providecommand \@url [1]{\endgroup\@href {#1}{\urlprefix }}%
\providecommand \urlprefix  [0]{URL }%
\providecommand \Eprint [0]{\href }%
\providecommand \doibase [0]{http://dx.doi.org/}%
\providecommand \selectlanguage [0]{\@gobble}%
\providecommand \bibinfo  [0]{\@secondoftwo}%
\providecommand \bibfield  [0]{\@secondoftwo}%
\providecommand \translation [1]{[#1]}%
\providecommand \BibitemOpen [0]{}%
\providecommand \bibitemStop [0]{}%
\providecommand \bibitemNoStop [0]{.\EOS\space}%
\providecommand \EOS [0]{\spacefactor3000\relax}%
\providecommand \BibitemShut  [1]{\csname bibitem#1\endcsname}%
\let\auto@bib@innerbib\@empty
\bibitem [{\citenamefont {{IceCube Collaboration}}(2013)}]{IceCube2013}%
  \BibitemOpen
  \bibfield  {author} {\bibinfo {author} {\bibnamefont {{IceCube
  Collaboration}}},\ }\bibfield  {title} {\enquote {\bibinfo {title} {{Evidence
  for High-Energy Extraterrestrial Neutrinos at the IceCube Detector}},}\
  }\href {\doibase 10.1126/science.1242856} {\bibfield  {journal} {\bibinfo
  {journal} {Science}\ }\textbf {\bibinfo {volume} {342}},\ \bibinfo {eid}
  {1242856} (\bibinfo {year} {2013})},\ \Eprint
  {http://arxiv.org/abs/1311.5238}{arXiv:1311.5238}\BibitemShut {NoStop}%
\bibitem [{\citenamefont {{Kadler}}\ \emph {{\it et~al.}}(2016)\citenamefont
  {{Kadler}}, \citenamefont {{Krau{\ss}}}, \citenamefont {{Mannheim}},
  \citenamefont {{Ojha}}, \citenamefont {{M{\"u}ller}}, \citenamefont
  {{Schulz}}, \citenamefont {{Anton}}, \citenamefont {{Baumgartner}},
  \citenamefont {{Beuchert}}, \citenamefont {{Buson}}, \citenamefont
  {{Carpenter}}, \citenamefont {{Eberl}}, \citenamefont {{Edwards}},
  \citenamefont {{Eisenacher Glawion}}, \citenamefont {{Els{\"a}sser}},
  \citenamefont {{Gehrels}}, \citenamefont {{Gr{\"a}fe}}, \citenamefont
  {{Gulyaev}}, \citenamefont {{Hase}}, \citenamefont {{Horiuchi}},
  \citenamefont {{James}}, \citenamefont {{Kappes}}, \citenamefont {{Kappes}},
  \citenamefont {{Katz}}, \citenamefont {{Kreikenbohm}}, \citenamefont
  {{Kreter}}, \citenamefont {{Kreykenbohm}}, \citenamefont {{Langejahn}},
  \citenamefont {{Leiter}}, \citenamefont {{Litzinger}}, \citenamefont
  {{Longo}}, \citenamefont {{Lovell}}, \citenamefont {{McEnery}}, \citenamefont
  {{Natusch}}, \citenamefont {{Phillips}}, \citenamefont {{Pl{\"o}tz}},
  \citenamefont {{Quick}}, \citenamefont {{Ros}}, \citenamefont {{Stecker}},
  \citenamefont {{Steinbring}}, \citenamefont {{Stevens}}, \citenamefont
  {{Thompson}}, \citenamefont {{Tr{\"u}stedt}}, \citenamefont {{Tzioumis}},
  \citenamefont {{Weston}}, \citenamefont {{Wilms}},\ and\ \citenamefont
  {{Zensus}}}]{Kadler2016BlazarNeutrino}%
  \BibitemOpen
  \bibfield  {author} {\bibinfo {author} {\bibfnamefont {M.}~\bibnamefont
  {{Kadler}}} {\it et~al.},\ }\bibfield  {title} {\enquote {\bibinfo {title}
  {{Coincidence of a high-fluence blazar outburst with a PeV-energy neutrino
  event}},}\ }\href {\doibase 10.1038/nphys3715} {\bibfield  {journal}
  {\bibinfo  {journal} {Nature Physics}\ }\textbf {\bibinfo {volume} {12}},\
  \bibinfo {pages} {807} (\bibinfo {year} {2016})},\ \Eprint
  {http://arxiv.org/abs/1602.02012}{arXiv:1602.02012}\BibitemShut {NoStop}%
\bibitem [{\citenamefont {{IceCube Collaboration}}\ \emph {{\it
  et~al.}}(2018{\natexlab{a}})\citenamefont {{IceCube Collaboration}},
  \citenamefont {{Aartsen}}, \citenamefont {{Ackermann}}, \citenamefont
  {{Adams}}, \citenamefont {{Aguilar}}, \citenamefont {{Ahlers}}, \citenamefont
  {{Ahrens}}, \citenamefont {{Al Samarai}}, \citenamefont {{Altmann}},
  \citenamefont {{Andeen}},\ and\ \citenamefont
  {et~al.}}]{IceCube2018MultiMessenger}%
  \BibitemOpen
  \bibfield  {author} {\bibinfo {author} {\bibnamefont {{IceCube
  Collaboration}}} {\it et~al.},\ }\bibfield  {title} {\enquote {\bibinfo
  {title} {{Multimessenger observations of a flaring blazar coincident with
  high-energy neutrino IceCube-170922A}},}\ }\href {\doibase
  10.1126/science.aat1378} {\bibfield  {journal} {\bibinfo  {journal}
  {Science}\ }\textbf {\bibinfo {volume} {361}},\ \bibinfo {eid} {eaat1378}
  (\bibinfo {year} {2018}{\natexlab{a}})}\BibitemShut {NoStop}%
\bibitem [{\citenamefont {{Liao}}\ \emph {{\it et~al.}}(2018)\citenamefont
  {{Liao}}, \citenamefont {{Xin}}, \citenamefont {{Liang}}, \citenamefont
  {{Guo}}, \citenamefont {{Li}}, \citenamefont {{He}}, \citenamefont {{Yuan}},\
  and\ \citenamefont {{Fan}}}]{Liao2018}%
  \BibitemOpen
  \bibfield  {author} {\bibinfo {author} {\bibfnamefont {N.-H.}\ \bibnamefont
  {{Liao}}}, \bibinfo {author} {\bibfnamefont {Y.-L.}\ \bibnamefont {{Xin}}},
  \bibinfo {author} {\bibfnamefont {Y.-F.}\ \bibnamefont {{Liang}}}, \bibinfo
  {author} {\bibfnamefont {X.-L.}\ \bibnamefont {{Guo}}}, \bibinfo {author}
  {\bibfnamefont {S.}~\bibnamefont {{Li}}}, \bibinfo {author} {\bibfnamefont
  {H.-N.}\ \bibnamefont {{He}}}, \bibinfo {author} {\bibfnamefont
  {Q.}~\bibnamefont {{Yuan}}}, and\ \bibinfo {author} {\bibfnamefont {Y.-Z.}\
  \bibnamefont {{Fan}}},\ }\bibfield  {title} {\enquote {\bibinfo {title}
  {{Active galactic nuclei with GeV activities and the PeV neutrino source
  candidate TXS 0506+056}},}\ }\href@noop {} {\bibfield  {journal} {\bibinfo
  {journal} {ArXiv e-prints}\ } (\bibinfo {year} {2018})},\ \Eprint
  {http://arxiv.org/abs/1807.05210}{arXiv:1807.05210}\BibitemShut {NoStop}%
\bibitem [{\citenamefont {{IceCube Collaboration}}\ \emph {{\it
  et~al.}}(2018{\natexlab{b}})\citenamefont {{IceCube Collaboration}},
  \citenamefont {{Aartsen}}, \citenamefont {{Ackermann}}, \citenamefont
  {{Adams}}, \citenamefont {{Aguilar}}, \citenamefont {{Ahlers}}, \citenamefont
  {{Ahrens}}, \citenamefont {{Samarai}}, \citenamefont {{Altmann}},
  \citenamefont {{Andeen}},\ and\ \citenamefont
  {et~al.}}]{IceCube2018neutrinoflare}%
  \BibitemOpen
  \bibfield  {author} {\bibinfo {author} {\bibnamefont {{IceCube
  Collaboration}}} {\it et~al.},\ }\bibfield  {title} {\enquote {\bibinfo
  {title} {{Neutrino emission from the direction of the blazar TXS 0506+056
  prior to the IceCube-170922A alert}},}\ }\href {\doibase
  10.1126/science.aat2890} {\bibfield  {journal} {\bibinfo  {journal}
  {Science}\ }\textbf {\bibinfo {volume} {361}},\ \bibinfo {pages} {147}
  (\bibinfo {year} {2018}{\natexlab{b}})},\ \Eprint
  {http://arxiv.org/abs/1807.08794}{arXiv:1807.08794}\BibitemShut {NoStop}%
\bibitem [{\citenamefont {{Protheroe}}(1997)}]{Protheroe1997Blazar}%
  \BibitemOpen
  \bibfield  {author} {\bibinfo {author} {\bibfnamefont {R.~J.}\ \bibnamefont
  {{Protheroe}}},\ }\bibfield  {title} {\enquote {\bibinfo {title} {{High
  Energy Neutrinos from Blazars}},}\ }in\ \href@noop {} {\emph {\bibinfo
  {booktitle} {IAU Colloq. 163}}},\ \bibinfo {series} {Astronomical Society of
  the Pacific Conference Series}, Vol.\ \bibinfo {volume} {121},\ \bibinfo
  {editor} {edited by\ \bibinfo {editor} {\bibfnamefont {D.~T.}\ \bibnamefont
  {{Wickramasinghe}}}, \bibinfo {editor} {\bibfnamefont {G.~V.}\ \bibnamefont
  {{Bicknell}}}, and\ \bibinfo {editor} {\bibfnamefont {L.}~\bibnamefont
  {{Ferrario}}}}\ (\bibinfo {year} {1997})\ p.\ \bibinfo {pages} {585},\
  \Eprint {http://arxiv.org/abs/astro-ph/9607165}{astro-ph/9607165}\BibitemShut
  {NoStop}%
\bibitem [{\citenamefont {{Dermer}}\ and\ \citenamefont
  {{Atoyan}}(2001)}]{Dermer2001Blazar}%
  \BibitemOpen
  \bibfield  {author} {\bibinfo {author} {\bibfnamefont {C.~D.}\ \bibnamefont
  {{Dermer}}} and\ \bibinfo {author} {\bibfnamefont {A.}~\bibnamefont
  {{Atoyan}}},\ }\bibfield  {title} {\enquote {\bibinfo {title} {{High-Energy
  Neutrino Production through Photopion Processes in Blazars}},}\ }\href@noop
  {} {\bibfield  {journal} {\bibinfo  {journal} {ArXiv Astrophysics e-prints}\
  } (\bibinfo {year} {2001})},\ \Eprint
  {http://arxiv.org/abs/astro-ph/0107200}{astro-ph/0107200}\BibitemShut
  {NoStop}%
\bibitem [{\citenamefont {{Gao}}\ \emph {{\it et~al.}}(2018)\citenamefont
  {{Gao}}, \citenamefont {{Fedynitch}}, \citenamefont {{Winter}},\ and\
  \citenamefont {{Pohl}}}]{Gao2018blazarneutrino}%
  \BibitemOpen
  \bibfield  {author} {\bibinfo {author} {\bibfnamefont {S.}~\bibnamefont
  {{Gao}}}, \bibinfo {author} {\bibfnamefont {A.}~\bibnamefont {{Fedynitch}}},
  \bibinfo {author} {\bibfnamefont {W.}~\bibnamefont {{Winter}}}, and\ \bibinfo
  {author} {\bibfnamefont {M.}~\bibnamefont {{Pohl}}},\ }\bibfield  {title}
  {\enquote {\bibinfo {title} {{Interpretation of the coincident observation of
  a high energy neutrino and a bright flare}},}\ }\href@noop {} {\bibfield
  {journal} {\bibinfo  {journal} {ArXiv e-prints}\ } (\bibinfo {year}
  {2018})},\ \Eprint
  {http://arxiv.org/abs/1807.04275}{arXiv:1807.04275}\BibitemShut {NoStop}%
\bibitem [{\citenamefont {{Murase}}\ \emph {{\it et~al.}}(2018)\citenamefont
  {{Murase}}, \citenamefont {{Oikonomou}},\ and\ \citenamefont
  {{Petropoulou}}}]{Murase2018blazarneutrino}%
  \BibitemOpen
  \bibfield  {author} {\bibinfo {author} {\bibfnamefont {K.}~\bibnamefont
  {{Murase}}}, \bibinfo {author} {\bibfnamefont {F.}~\bibnamefont
  {{Oikonomou}}}, and\ \bibinfo {author} {\bibfnamefont {M.}~\bibnamefont
  {{Petropoulou}}},\ }\bibfield  {title} {\enquote {\bibinfo {title} {{Blazar
  Flares as an Origin of High-Energy Cosmic Neutrinos?}}}\ }\href@noop {}
  {\bibfield  {journal} {\bibinfo  {journal} {ArXiv e-prints}\ } (\bibinfo
  {year} {2018})},\ \Eprint
  {http://arxiv.org/abs/1807.04748}{arXiv:1807.04748}\BibitemShut {NoStop}%
\bibitem [{\citenamefont {{Righi}}\ \emph {{\it et~al.}}(2018)\citenamefont
  {{Righi}}, \citenamefont {{Tavecchio}},\ and\ \citenamefont
  {{Pacciani}}}]{Righi2018blazarneutrino}%
  \BibitemOpen
  \bibfield  {author} {\bibinfo {author} {\bibfnamefont {C.}~\bibnamefont
  {{Righi}}}, \bibinfo {author} {\bibfnamefont {F.}~\bibnamefont
  {{Tavecchio}}}, and\ \bibinfo {author} {\bibfnamefont {L.}~\bibnamefont
  {{Pacciani}}},\ }\bibfield  {title} {\enquote {\bibinfo {title} {{A
  multiwavelength view of BL Lacs neutrino candidates}},}\ }\href@noop {}
  {\bibfield  {journal} {\bibinfo  {journal} {ArXiv e-prints}\ } (\bibinfo
  {year} {2018})},\ \Eprint
  {http://arxiv.org/abs/1807.04299}{arXiv:1807.04299}\BibitemShut {NoStop}%
\bibitem [{\citenamefont {{Ansoldi}}\ \emph {{\it et~al.}}(2018)\citenamefont
  {{Ansoldi}}, \citenamefont {{Antonelli}}, \citenamefont {{Arcaro}},
  \citenamefont {{Baack}}, \citenamefont {{Babi{\'c}}}, \citenamefont
  {{Banerjee}}, \citenamefont {{Bangale}}, \citenamefont {{Barres de Almeida}},
  \citenamefont {{Barrio}},\ and\ \citenamefont {{Becerra
  Gonz{\'a}lez}}}]{Ansoldi2018blazarneutrino}%
  \BibitemOpen
  \bibfield  {author} {\bibinfo {author} {\bibfnamefont {S.}~\bibnamefont
  {{Ansoldi}}}, \bibinfo {author} {\bibfnamefont {L.~A.}\ \bibnamefont
  {{Antonelli}}}, \bibinfo {author} {\bibfnamefont {C.}~\bibnamefont
  {{Arcaro}}}, \bibinfo {author} {\bibfnamefont {D.}~\bibnamefont {{Baack}}},
  \bibinfo {author} {\bibfnamefont {A.}~\bibnamefont {{Babi{\'c}}}}, \bibinfo
  {author} {\bibfnamefont {B.}~\bibnamefont {{Banerjee}}}, \bibinfo {author}
  {\bibfnamefont {P.}~\bibnamefont {{Bangale}}}, \bibinfo {author}
  {\bibfnamefont {U.}~\bibnamefont {{Barres de Almeida}}}, \bibinfo {author}
  {\bibfnamefont {J.~A.}\ \bibnamefont {{Barrio}}}, and\ \bibinfo {author}
  {\bibfnamefont {J.~e.~a.}\ \bibnamefont {{Becerra Gonz{\'a}lez}}},\
  }\bibfield  {title} {\enquote {\bibinfo {title} {{The blazar TXS 0506+056
  associated with a high-energy neutrino: insights into extragalactic jets and
  cosmic ray acceleration}},}\ }\href@noop {} {\bibfield  {journal} {\bibinfo
  {journal} {ArXiv e-prints}\ } (\bibinfo {year} {2018})},\ \Eprint
  {http://arxiv.org/abs/1807.04300}{arXiv:1807.04300}\BibitemShut {NoStop}%
\bibitem [{\citenamefont {{Keivani}}\ \emph {{\it et~al.}}(2018)\citenamefont
  {{Keivani}}, \citenamefont {{Murase}}, \citenamefont {{Petropoulou}},
  \citenamefont {{Fox}}, \citenamefont {{Cenko}}, \citenamefont {{Chaty}},
  \citenamefont {{Coleiro}}, \citenamefont {{DeLaunay}}, \citenamefont
  {{Dimitrakoudis}}, \citenamefont {{Evans}}, \citenamefont {{Kennea}},
  \citenamefont {{Marshall}}, \citenamefont {{Mastichiadis}}, \citenamefont
  {{Osborne}}, \citenamefont {{Santander}}, \citenamefont {{Tohuvavohu}},\ and\
  \citenamefont {{Turley}}}]{Keivani2018blazarneutrino}%
  \BibitemOpen
  \bibfield  {author} {\bibinfo {author} {\bibfnamefont {A.}~\bibnamefont
  {{Keivani}}} {\it et~al.},\ }\bibfield  {title} {\enquote {\bibinfo {title}
  {{A Multimessenger Picture of the Flaring Blazar TXS 0506+056: implications
  for High-Energy Neutrino Emission and Cosmic Ray Acceleration}},}\
  }\href@noop {} {\bibfield  {journal} {\bibinfo  {journal} {ArXiv e-prints}\ }
  (\bibinfo {year} {2018})},\ \Eprint
  {http://arxiv.org/abs/1807.04537}{arXiv:1807.04537}\BibitemShut {NoStop}%
\bibitem [{\citenamefont {{Rodrigues}}\ \emph {{\it et~al.}}(2018)\citenamefont
  {{Rodrigues}}, \citenamefont {{Fedynitch}}, \citenamefont {{Gao}},
  \citenamefont {{Boncioli}},\ and\ \citenamefont
  {{Winter}}}]{Rodrigures2018blazarneutrino}%
  \BibitemOpen
  \bibfield  {author} {\bibinfo {author} {\bibfnamefont {X.}~\bibnamefont
  {{Rodrigues}}}, \bibinfo {author} {\bibfnamefont {A.}~\bibnamefont
  {{Fedynitch}}}, \bibinfo {author} {\bibfnamefont {S.}~\bibnamefont {{Gao}}},
  \bibinfo {author} {\bibfnamefont {D.}~\bibnamefont {{Boncioli}}}, and\
  \bibinfo {author} {\bibfnamefont {W.}~\bibnamefont {{Winter}}},\ }\bibfield
  {title} {\enquote {\bibinfo {title} {{Neutrinos and Ultra-high-energy
  Cosmic-ray Nuclei from Blazars}},}\ }\href {\doibase
  10.3847/1538-4357/aaa7ee} {\bibfield  {journal} {\bibinfo  {journal}
  {Astrophys. J.}\ }\textbf {\bibinfo {volume} {854}},\ \bibinfo {eid} {54}
  (\bibinfo {year} {2018})},\ \Eprint
  {http://arxiv.org/abs/1711.02091}{arXiv:1711.02091}\BibitemShut {NoStop}%
\bibitem [{\citenamefont {{Liu}}\ \emph {{\it et~al.}}(2018)\citenamefont
  {{Liu}}, \citenamefont {{Wang}}, \citenamefont {{Xue}}, \citenamefont
  {{Taylor}}, \citenamefont {{Wang}}, \citenamefont {{Li}},\ and\ \citenamefont
  {{Yan}}}]{Liu2018blazarneutrino}%
  \BibitemOpen
  \bibfield  {author} {\bibinfo {author} {\bibfnamefont {R.-Y.}\ \bibnamefont
  {{Liu}}}, \bibinfo {author} {\bibfnamefont {K.}~\bibnamefont {{Wang}}},
  \bibinfo {author} {\bibfnamefont {R.}~\bibnamefont {{Xue}}}, \bibinfo
  {author} {\bibfnamefont {A.~M.}\ \bibnamefont {{Taylor}}}, \bibinfo {author}
  {\bibfnamefont {X.-Y.}\ \bibnamefont {{Wang}}}, \bibinfo {author}
  {\bibfnamefont {Z.}~\bibnamefont {{Li}}}, and\ \bibinfo {author}
  {\bibfnamefont {H.}~\bibnamefont {{Yan}}},\ }\bibfield  {title} {\enquote
  {\bibinfo {title} {{A hadronuclear interpretation of a high-energy neutrino
  event coincident with a blazar flare}},}\ }\href@noop {} {\bibfield
  {journal} {\bibinfo  {journal} {ArXiv e-prints}\ } (\bibinfo {year}
  {2018})},\ \Eprint
  {http://arxiv.org/abs/1807.05113}{arXiv:1807.05113}\BibitemShut {NoStop}%
\bibitem [{\citenamefont {{Padovani}}\ \emph {{\it et~al.}}(2018)\citenamefont
  {{Padovani}}, \citenamefont {{Giommi}}, \citenamefont {{Resconi}},
  \citenamefont {{Glauch}}, \citenamefont {{Arsioli}}, \citenamefont
  {{Sahakyan}},\ and\ \citenamefont {{Huber}}}]{Padovani2018}%
  \BibitemOpen
  \bibfield  {author} {\bibinfo {author} {\bibfnamefont {P.}~\bibnamefont
  {{Padovani}}}, \bibinfo {author} {\bibfnamefont {P.}~\bibnamefont
  {{Giommi}}}, \bibinfo {author} {\bibfnamefont {E.}~\bibnamefont {{Resconi}}},
  \bibinfo {author} {\bibfnamefont {T.}~\bibnamefont {{Glauch}}}, \bibinfo
  {author} {\bibfnamefont {B.}~\bibnamefont {{Arsioli}}}, \bibinfo {author}
  {\bibfnamefont {N.}~\bibnamefont {{Sahakyan}}}, and\ \bibinfo {author}
  {\bibfnamefont {M.}~\bibnamefont {{Huber}}},\ }\bibfield  {title} {\enquote
  {\bibinfo {title} {{Dissecting the region around IceCube-170922A: the blazar
  TXS 0506+056 as the first cosmic neutrino source}},}\ }\href {\doibase
  10.1093/mnras/sty1852} {\bibfield  {journal} {\bibinfo  {journal} {\mnras}\
  }\textbf {\bibinfo {volume} {480}},\ \bibinfo {pages} {192} (\bibinfo {year}
  {2018})},\ \Eprint
  {http://arxiv.org/abs/1807.04461}{arXiv:1807.04461}\BibitemShut {NoStop}%
\bibitem [{\citenamefont {{Johnston}}\ \emph {{\it et~al.}}(1995)\citenamefont
  {{Johnston}}, \citenamefont {{Fey}}, \citenamefont {{Zacharias}},
  \citenamefont {{Russell}}, \citenamefont {{Ma}}, \citenamefont {{de Vegt}},
  \citenamefont {{Reynolds}}, \citenamefont {{Jauncey}}, \citenamefont
  {{Archinal}}, \citenamefont {{Carter}}, \citenamefont {{Corbin}},
  \citenamefont {{Eubanks}}, \citenamefont {{Florkowski}}, \citenamefont
  {{Hall}}, \citenamefont {{McCarthy}}, \citenamefont {{McCulloch}},
  \citenamefont {{King}}, \citenamefont {{Nicolson}},\ and\ \citenamefont
  {{Shaffer}}}]{Johnston1995}%
  \BibitemOpen
  \bibfield  {author} {\bibinfo {author} {\bibfnamefont {K.~J.}\ \bibnamefont
  {{Johnston}}} {\it et~al.},\ }\bibfield  {title} {\enquote {\bibinfo {title}
  {{A Radio Reference Frame}},}\ }\href {\doibase 10.1086/117571} {\bibfield
  {journal} {\bibinfo  {journal} {\aj}\ }\textbf {\bibinfo {volume} {110}},\
  \bibinfo {pages} {880} (\bibinfo {year} {1995})}\BibitemShut {NoStop}%
\bibitem [{\citenamefont {{Atwood}}\ \emph {{\it et~al.}}(2009)\citenamefont
  {{Atwood}}, \citenamefont {{Abdo}}, \citenamefont {{Ackermann}},
  \citenamefont {{Althouse}}, \citenamefont {{Anderson}}, \citenamefont
  {{Axelsson}}, \citenamefont {{Baldini}}, \citenamefont {{Ballet}},
  \citenamefont {{Band}}, \citenamefont {{Barbiellini}},\ and\ \citenamefont
  {et~al.}}]{Atwood2009}%
  \BibitemOpen
  \bibfield  {author} {\bibinfo {author} {\bibfnamefont {W.~B.}\ \bibnamefont
  {{Atwood}}} {\it et~al.},\ }\bibfield  {title} {\enquote {\bibinfo {title}
  {{The Large Area Telescope on the Fermi Gamma-Ray Space Telescope
  Mission}},}\ }\href {\doibase 10.1088/0004-637X/697/2/1071} {\bibfield
  {journal} {\bibinfo  {journal} {Astrophys. J.}\ }\textbf {\bibinfo {volume}
  {697}},\ \bibinfo {pages} {1071} (\bibinfo {year} {2009})},\ \Eprint
  {http://arxiv.org/abs/0902.1089}{arXiv:0902.1089}\BibitemShut {NoStop}%
\bibitem [{\citenamefont {{Atwood}}\ \emph {{\it et~al.}}(2013)\citenamefont
  {{Atwood}}, \citenamefont {{Albert}}, \citenamefont {{Baldini}},
  \citenamefont {{Tinivella}}, \citenamefont {{Bregeon}}, \citenamefont
  {{Pesce-Rollins}}, \citenamefont {{Sgr{\`o}}}, \citenamefont {{Bruel}},
  \citenamefont {{Charles}}, \citenamefont {{Drlica-Wagner}}, \citenamefont
  {{Franckowiak}}, \citenamefont {{Jogler}}, \citenamefont {{Rochester}},
  \citenamefont {{Usher}}, \citenamefont {{Wood}}, \citenamefont
  {{Cohen-Tanugi}},\ and\ \citenamefont {{S.~Zimmer for the Fermi-LAT
  Collaboration}}}]{Atwood2013}%
  \BibitemOpen
  \bibfield  {author} {\bibinfo {author} {\bibfnamefont {W.}~\bibnamefont
  {{Atwood}}} {\it et~al.},\ }\bibfield  {title} {\enquote {\bibinfo {title}
  {{Pass 8: Toward the Full Realization of the Fermi-LAT Scientific
  Potential}},}\ }\href@noop {} {\bibfield  {journal} {\bibinfo  {journal}
  {ArXiv e-prints}\ } (\bibinfo {year} {2013})},\ \Eprint
  {http://arxiv.org/abs/1303.3514}{arXiv:1303.3514}\BibitemShut {NoStop}%
\bibitem [{\citenamefont {{Gaia Collaboration}}(2018)}]{Gaia2018}%
  \BibitemOpen
  \bibfield  {author} {\bibinfo {author} {\bibnamefont {{Gaia
  Collaboration}}},\ }\bibfield  {title} {\enquote {\bibinfo {title} {{VizieR
  Online Data Catalog: Gaia DR2 (Gaia Collaboration, 2018)}},}\ }\href@noop {}
  {\bibfield  {journal} {\bibinfo  {journal} {VizieR Online Data Catalog}\
  }\textbf {\bibinfo {volume} {1345}} (\bibinfo {year} {2018})}\BibitemShut
  {NoStop}%
\bibitem [{\citenamefont {{Drinkwater}}\ \emph {{\it
  et~al.}}(1997)\citenamefont {{Drinkwater}}, \citenamefont {{Webster}},
  \citenamefont {{Francis}}, \citenamefont {{Condon}}, \citenamefont
  {{Ellison}}, \citenamefont {{Jauncey}}, \citenamefont {{Lovell}},
  \citenamefont {{Peterson}},\ and\ \citenamefont {{Savage}}}]{Drinkwater1997}%
  \BibitemOpen
  \bibfield  {author} {\bibinfo {author} {\bibfnamefont {M.~J.}\ \bibnamefont
  {{Drinkwater}}}, \bibinfo {author} {\bibfnamefont {R.~L.}\ \bibnamefont
  {{Webster}}}, \bibinfo {author} {\bibfnamefont {P.~J.}\ \bibnamefont
  {{Francis}}}, \bibinfo {author} {\bibfnamefont {J.~J.}\ \bibnamefont
  {{Condon}}}, \bibinfo {author} {\bibfnamefont {S.~L.}\ \bibnamefont
  {{Ellison}}}, \bibinfo {author} {\bibfnamefont {D.~L.}\ \bibnamefont
  {{Jauncey}}}, \bibinfo {author} {\bibfnamefont {J.}~\bibnamefont {{Lovell}}},
  \bibinfo {author} {\bibfnamefont {B.~A.}\ \bibnamefont {{Peterson}}}, and\
  \bibinfo {author} {\bibfnamefont {A.}~\bibnamefont {{Savage}}},\ }\bibfield
  {title} {\enquote {\bibinfo {title} {{The Parkes Half-Jansky Flat-Spectrum
  Sample}},}\ }\href {\doibase 10.1093/mnras/284.1.85} {\bibfield  {journal}
  {\bibinfo  {journal} {\mnras}\ }\textbf {\bibinfo {volume} {284}},\ \bibinfo
  {pages} {85} (\bibinfo {year} {1997})},\ \Eprint
  {http://arxiv.org/abs/astro-ph/9609019}{astro-ph/9609019}\BibitemShut
  {NoStop}%
\bibitem [{\citenamefont {{Poutanen}}\ and\ \citenamefont
  {{Stern}}(2010)}]{Poutanen2010}%
  \BibitemOpen
  \bibfield  {author} {\bibinfo {author} {\bibfnamefont {J.}~\bibnamefont
  {{Poutanen}}} and\ \bibinfo {author} {\bibfnamefont {B.}~\bibnamefont
  {{Stern}}},\ }\bibfield  {title} {\enquote {\bibinfo {title} {{GeV Breaks in
  Blazars as a Result of Gamma-ray Absorption Within the Broad-line Region}},}\
  }\href {\doibase 10.1088/2041-8205/717/2/L118} {\bibfield  {journal}
  {\bibinfo  {journal} {\apjl}\ }\textbf {\bibinfo {volume} {717}},\ \bibinfo
  {pages} {L118} (\bibinfo {year} {2010})},\ \Eprint
  {http://arxiv.org/abs/1005.3792}{arXiv:1005.3792}\BibitemShut {NoStop}%
\bibitem [{\citenamefont {{Stern}}\ and\ \citenamefont
  {{Poutanen}}(2014)}]{Stern2014nH}%
  \BibitemOpen
  \bibfield  {author} {\bibinfo {author} {\bibfnamefont {B.~E.}\ \bibnamefont
  {{Stern}}} and\ \bibinfo {author} {\bibfnamefont {J.}~\bibnamefont
  {{Poutanen}}},\ }\bibfield  {title} {\enquote {\bibinfo {title} {{The Mystery
  of Spectral Breaks: Lyman Continuum Absorption by Photon-Photon Pair
  Production in the Fermi GeV Spectra of Bright Blazars}},}\ }\href {\doibase
  10.1088/0004-637X/794/1/8} {\bibfield  {journal} {\bibinfo  {journal}
  {Astrophys. J.}\ }\textbf {\bibinfo {volume} {794}},\ \bibinfo {eid} {8}
  (\bibinfo {year} {2014})},\ \Eprint
  {http://arxiv.org/abs/1408.0793}{arXiv:1408.0793}\BibitemShut {NoStop}%
\bibitem [{\citenamefont {{He}}\ \emph {{\it et~al.}}(2018)\citenamefont
  {{He}}, \citenamefont {{Inoue}}, \citenamefont {{Inoue}},\ and\ \citenamefont
  {{Liang}}}]{he2018}%
  \BibitemOpen
  \bibfield  {author} {\bibinfo {author} {\bibfnamefont {H.-N.}\ \bibnamefont
  {{He}}}, \bibinfo {author} {\bibfnamefont {Y.}~\bibnamefont {{Inoue}}},
  \bibinfo {author} {\bibfnamefont {S.}~\bibnamefont {{Inoue}}}, and\ \bibinfo
  {author} {\bibfnamefont {Y.-F.}\ \bibnamefont {{Liang}}},\ }\bibfield
  {title} {\enquote {\bibinfo {title} {{High-energy neutrino flare from
  cloud-jet interaction in the blazar PKS 0502+049}},}\ }\href@noop {}
  {\bibfield  {journal} {\bibinfo  {journal} {ArXiv e-prints}\ } (\bibinfo
  {year} {2018})},\ \Eprint
  {http://arxiv.org/abs/1808.04330}{arXiv:1808.04330}\BibitemShut {NoStop}%
\bibitem [{\citenamefont {{Aartsen}}\ \emph {{\it
  et~al.}}(2017{\natexlab{a}})\citenamefont {{Aartsen}}, \citenamefont
  {{Abraham}}, \citenamefont {{Ackermann}}, \citenamefont {{Adams}},
  \citenamefont {{Aguilar}}, \citenamefont {{Ahlers}}, \citenamefont
  {{Ahrens}}, \citenamefont {{Altmann}}, \citenamefont {{Andeen}},
  \citenamefont {{Anderson}},\ and\ \citenamefont
  {et~al.}}]{IceCube20177years}%
  \BibitemOpen
  \bibfield  {author} {\bibinfo {author} {\bibfnamefont {M.~G.}\ \bibnamefont
  {{Aartsen}}} {\it et~al.},\ }\bibfield  {title} {\enquote {\bibinfo {title}
  {{All-sky Search for Time-integrated Neutrino Emission from Astrophysical
  Sources with 7 yr of IceCube Data}},}\ }\href {\doibase
  10.3847/1538-4357/835/2/151} {\bibfield  {journal} {\bibinfo  {journal}
  {Astrophys. J.}\ }\textbf {\bibinfo {volume} {835}},\ \bibinfo {eid} {151}
  (\bibinfo {year} {2017}{\natexlab{a}})},\ \Eprint
  {http://arxiv.org/abs/1609.04981}{arXiv:1609.04981}\BibitemShut {NoStop}%
\bibitem [{\citenamefont {{Kelner}}\ \emph {{\it et~al.}}(2006)\citenamefont
  {{Kelner}}, \citenamefont {{Aharonian}},\ and\ \citenamefont
  {{Bugayov}}}]{Kelner2006pp}%
  \BibitemOpen
  \bibfield  {author} {\bibinfo {author} {\bibfnamefont {S.~R.}\ \bibnamefont
  {{Kelner}}}, \bibinfo {author} {\bibfnamefont {F.~A.}\ \bibnamefont
  {{Aharonian}}}, and\ \bibinfo {author} {\bibfnamefont {V.~V.}\ \bibnamefont
  {{Bugayov}}},\ }\bibfield  {title} {\enquote {\bibinfo {title} {{Energy
  spectra of gamma rays, electrons, and neutrinos produced at proton-proton
  interactions in the very high energy regime}},}\ }\href {\doibase
  10.1103/PhysRevD.74.034018} {\bibfield  {journal} {\bibinfo  {journal}
  {\prd}\ }\textbf {\bibinfo {volume} {74}},\ \bibinfo {eid} {034018} (\bibinfo
  {year} {2006})},\ \Eprint
  {http://arxiv.org/abs/astro-ph/0606058}{astro-ph/0606058}\BibitemShut
  {NoStop}%
\bibitem [{\citenamefont {{Paiano}}\ \emph {{\it et~al.}}(2018)\citenamefont
  {{Paiano}}, \citenamefont {{Falomo}}, \citenamefont {{Treves}},\ and\
  \citenamefont {{Scarpa}}}]{Paiano2018redshift}%
  \BibitemOpen
  \bibfield  {author} {\bibinfo {author} {\bibfnamefont {S.}~\bibnamefont
  {{Paiano}}}, \bibinfo {author} {\bibfnamefont {R.}~\bibnamefont {{Falomo}}},
  \bibinfo {author} {\bibfnamefont {A.}~\bibnamefont {{Treves}}}, and\ \bibinfo
  {author} {\bibfnamefont {R.}~\bibnamefont {{Scarpa}}},\ }\bibfield  {title}
  {\enquote {\bibinfo {title} {{The Redshift of the BL Lac Object TXS
  0506+056}},}\ }\href {\doibase 10.3847/2041-8213/aaad5e} {\bibfield
  {journal} {\bibinfo  {journal} {\apjl}\ }\textbf {\bibinfo {volume} {854}},\
  \bibinfo {eid} {L32} (\bibinfo {year} {2018})},\ \Eprint
  {http://arxiv.org/abs/1802.01939}{arXiv:1802.01939}\BibitemShut {NoStop}%
\bibitem [{\citenamefont {{Kochanek}}\ \emph {{\it et~al.}}(2017)\citenamefont
  {{Kochanek}}, \citenamefont {{Shappee}}, \citenamefont {{Stanek}},
  \citenamefont {{Holoien}}, \citenamefont {{Thompson}}, \citenamefont
  {{Prieto}}, \citenamefont {{Dong}}, \citenamefont {{Shields}}, \citenamefont
  {{Will}}, \citenamefont {{Britt}}, \citenamefont {{Perzanowski}},\ and\
  \citenamefont {{Pojma{\'n}ski}}}]{Kochanek2017ASAS-SN}%
  \BibitemOpen
  \bibfield  {author} {\bibinfo {author} {\bibfnamefont {C.~S.}\ \bibnamefont
  {{Kochanek}}} {\it et~al.},\ }\bibfield  {title} {\enquote {\bibinfo {title}
  {{The All-Sky Automated Survey for Supernovae (ASAS-SN) Light Curve Server
  v1.0}},}\ }\href {\doibase 10.1088/1538-3873/aa80d9} {\bibfield  {journal}
  {\bibinfo  {journal} {\pasp}\ }\textbf {\bibinfo {volume} {129}},\ \bibinfo
  {pages} {104502} (\bibinfo {year} {2017})},\ \Eprint
  {http://arxiv.org/abs/1706.07060}{arXiv:1706.07060}\BibitemShut {NoStop}%
\bibitem [{\citenamefont {{Braun}}\ \emph {{\it et~al.}}(2008)\citenamefont
  {{Braun}}, \citenamefont {{Dumm}}, \citenamefont {{De Palma}}, \citenamefont
  {{Finley}}, \citenamefont {{Karle}},\ and\ \citenamefont
  {{Montaruli}}}]{Braun2008pointsourceneutrino}%
  \BibitemOpen
  \bibfield  {author} {\bibinfo {author} {\bibfnamefont {J.}~\bibnamefont
  {{Braun}}}, \bibinfo {author} {\bibfnamefont {J.}~\bibnamefont {{Dumm}}},
  \bibinfo {author} {\bibfnamefont {F.}~\bibnamefont {{De Palma}}}, \bibinfo
  {author} {\bibfnamefont {C.}~\bibnamefont {{Finley}}}, \bibinfo {author}
  {\bibfnamefont {A.}~\bibnamefont {{Karle}}}, and\ \bibinfo {author}
  {\bibfnamefont {T.}~\bibnamefont {{Montaruli}}},\ }\bibfield  {title}
  {\enquote {\bibinfo {title} {{Methods for point source analysis in high
  energy neutrino telescopes}},}\ }\href {\doibase
  10.1016/j.astropartphys.2008.02.007} {\bibfield  {journal} {\bibinfo
  {journal} {Astroparticle Physics}\ }\textbf {\bibinfo {volume} {29}},\
  \bibinfo {pages} {299} (\bibinfo {year} {2008})},\ \Eprint
  {http://arxiv.org/abs/0801.1604}{arXiv:0801.1604}\BibitemShut {NoStop}%
\bibitem [{\citenamefont {{Braun}}\ \emph {{\it et~al.}}(2010)\citenamefont
  {{Braun}}, \citenamefont {{Baker}}, \citenamefont {{Dumm}}, \citenamefont
  {{Finley}}, \citenamefont {{Karle}},\ and\ \citenamefont
  {{Montaruli}}}]{Braun2010Time-dependent}%
  \BibitemOpen
  \bibfield  {author} {\bibinfo {author} {\bibfnamefont {J.}~\bibnamefont
  {{Braun}}}, \bibinfo {author} {\bibfnamefont {M.}~\bibnamefont {{Baker}}},
  \bibinfo {author} {\bibfnamefont {J.}~\bibnamefont {{Dumm}}}, \bibinfo
  {author} {\bibfnamefont {C.}~\bibnamefont {{Finley}}}, \bibinfo {author}
  {\bibfnamefont {A.}~\bibnamefont {{Karle}}}, and\ \bibinfo {author}
  {\bibfnamefont {T.}~\bibnamefont {{Montaruli}}},\ }\bibfield  {title}
  {\enquote {\bibinfo {title} {{Time-dependent point source search methods in
  high energy neutrino astronomy}},}\ }\href {\doibase
  10.1016/j.astropartphys.2010.01.005} {\bibfield  {journal} {\bibinfo
  {journal} {Astroparticle Physics}\ }\textbf {\bibinfo {volume} {33}},\
  \bibinfo {pages} {175} (\bibinfo {year} {2010})},\ \Eprint
  {http://arxiv.org/abs/0912.1572}{arXiv:0912.1572}\BibitemShut {NoStop}%
\bibitem [{\citenamefont {{Aartsen}}\ \emph {{\it et~al.}}(2015)\citenamefont
  {{Aartsen}}, \citenamefont {{Ackermann}}, \citenamefont {{Adams}},
  \citenamefont {{Aguilar}}, \citenamefont {{Ahlers}}, \citenamefont
  {{Ahrens}}, \citenamefont {{Altmann}}, \citenamefont {{Anderson}},
  \citenamefont {{Archinger}}, \citenamefont {{Arguelles}},\ and\ \citenamefont
  {et~al.}}]{IceCube2015timedependent}%
  \BibitemOpen
  \bibfield  {author} {\bibinfo {author} {\bibfnamefont {M.~G.}\ \bibnamefont
  {{Aartsen}}} {\it et~al.},\ }\bibfield  {title} {\enquote {\bibinfo {title}
  {{Searches for Time-dependent Neutrino Sources with IceCube Data from 2008 to
  2012}},}\ }\href {\doibase 10.1088/0004-637X/807/1/46} {\bibfield  {journal}
  {\bibinfo  {journal} {Astrophys. J.}\ }\textbf {\bibinfo {volume} {807}},\
  \bibinfo {eid} {46} (\bibinfo {year} {2015})},\ \Eprint
  {http://arxiv.org/abs/1503.00598}{arXiv:1503.00598}\BibitemShut {NoStop}%
\bibitem [{\citenamefont {{Aartsen}}\ \emph {{\it
  et~al.}}(2017{\natexlab{b}})\citenamefont {{Aartsen}}, \citenamefont
  {{Abraham}}, \citenamefont {{Ackermann}}, \citenamefont {{Adams}},
  \citenamefont {{Aguilar}}, \citenamefont {{Ahlers}}, \citenamefont
  {{Ahrens}}, \citenamefont {{Altmann}}, \citenamefont {{Andeen}},
  \citenamefont {{Anderson}},\ and\ \citenamefont
  {et~al.}}]{IceCube2017timeintegrated}%
  \BibitemOpen
  \bibfield  {author} {\bibinfo {author} {\bibfnamefont {M.~G.}\ \bibnamefont
  {{Aartsen}}} {\it et~al.},\ }\bibfield  {title} {\enquote {\bibinfo {title}
  {{All-sky Search for Time-integrated Neutrino Emission from Astrophysical
  Sources with 7 yr of IceCube Data}},}\ }\href {\doibase
  10.3847/1538-4357/835/2/151} {\bibfield  {journal} {\bibinfo  {journal}
  {Astrophys. J.}\ }\textbf {\bibinfo {volume} {835}},\ \bibinfo {eid} {151}
  (\bibinfo {year} {2017}{\natexlab{b}})},\ \Eprint
  {http://arxiv.org/abs/1609.04981}{arXiv:1609.04981}\BibitemShut {NoStop}%
\end{thebibliography}%
\end{document}